\documentclass[draftclsnofoot, 12pt, onecolumn]{IEEEtran}
\usepackage{bbm,amssymb,amsmath,amsthm,amsbsy,mathrsfs}
\usepackage{graphicx,epstopdf,epsfig,psfrag,url,cite,color,multirow,float,makecell}
\usepackage[small,bf]{caption}
\usepackage[font=footnotesize]{subfig}
\usepackage{algorithm,algorithmic}
\usepackage[T1]{fontenc}
\usepackage[latin9]{inputenc}
\usepackage{fixltx2e}
\usepackage{array,multirow,graphicx}
\usepackage{color}
\usepackage{cite}
\usepackage[x11names,dvipsnames,table]{xcolor} 
\usepackage{colortbl}
\usepackage{lipsum}
\definecolor{mygray}{gray}{0.6}
\definecolor{myblue}{rgb}{0.8,0.85,1}
\usepackage{array}
\newcolumntype{L}[1]{>{\raggedright\let\newline\\\arraybackslash\hspace{0pt}}m{#1}}
\newcolumntype{C}[1]{>{\centering\let\newline\\\arraybackslash\hspace{0pt}}m{#1}}
\newcolumntype{R}[1]{>{\raggedleft\let\newline\\\arraybackslash\hspace{0pt}}m{#1}}

\usepackage{array, tabularx, boldline}
\usepackage{eurosym}
\usepackage{amstext} 
\usepackage{cellspace}

\newtheoremstyle{mythm}
{\topsep}   
{\topsep}   
{\itshape}      
{0pt}       
{\bfseries} 
{:}         
{5pt plus 1pt minus 1pt}    
{\thmname{#1}\thmnumber{ #2}\thmnote{ (#3)}}
\theoremstyle{mythm}

\begin{document}

\title{Deep Reinforcement Learning for Backscatter-Aided Data Offloading in Mobile Edge Computing}

\author{
\IEEEauthorblockN{Shimin Gong, Yutong Xie, Jing Xu, Dusit Niyato, and Ying-Chang Liang}
\thanks{Shimin Gong is with the School of Intelligent Systems Engineering, Sun Yat-sen University, China, email: gongshm5@mail.sysu.edu.cn. Yutong Xie is with Shenzhen Institutes of Advanced Technology, Chinese Academy of Sciences, email: yt.xie@siat.ac.cn. Jing Xu is with the School of Electronic Information and Communications, Huazhong University of Science and Technology, China, email: xujing@hust.edu.cn. Dusit Niyato is with the School of Computer Science and Engineering, Nanyang Technological University, Singapore, email: dniyato@ntu.edu.sg. Ying-Chang Liang is with the Center for Intelligent Networking and Communications (CINC), University of Electronic Science and Technology of China, China, email: liangyc@ieee.org.
}
}

\maketitle
\thispagestyle{empty}

\begin{abstract}
Wireless network optimization has been becoming very challenging as the problem size and complexity increase tremendously, due to close couplings among network entities with heterogeneous service and resource requirements. By continuously interacting with the environment, deep reinforcement learning (DRL) provides a mechanism for different network entities to build knowledge and make autonomous decisions to improve network performance. In this article, we first review typical DRL approaches and recent enhancements. We then discuss the applications of DRL for mobile edge computing (MEC), which can be used for the low-power IoT devices, e.g., wireless sensors in healthcare monitoring, to offload computation workload to nearby MEC servers. To balance power consumption in offloading and computation, we propose a novel hybrid offloading model that exploits the complement operations of RF communications and low-power backscatter communications. The DRL framework is then customized to optimize the transmission scheduling and workload allocation in two communications technologies, which is shown to enhance the offloading performance significantly compared with existing schemes.
\end{abstract}

\newpage
\section{Introduction}
Modern wireless networks have to embrace the upsurge of traffic demands and diverse quality provisioning requirements. This requires a strategic shift in the network design that utilizes sophisticated wireless technologies in a more decentralized, ad-hoc, and diverse environment. As such, the network design problems become very challenging as the dimensionality and complexity rapidly increase, e.g., due to couplings among different network entities. Recently, deep reinforcement learning (DRL) has been developed as a breakthrough technology to learn the optimal control strategy in a dynamic network environment by continuously interacting with it~\cite{luong18}. DRL integrates deep neural networks (DNNs) with the conventional reinforcement learning algorithms for autonomous decision making. It becomes capable of solving high dimensional, non-convex, and model-free network control problems, e.g., channel access and resource allocation in mobile edge computing (MEC)~\cite{yan18}. These are very difficult to handle by conventional techniques such as convex optimization, dynamic and stochastic programming, due to imprecise modeling, uncertain system dynamics, and huge variable spaces. Hence, the application of DRL in wireless networks is envisioned to revolutionize the network optimization paradigm.

In this article, we first provide an overview of the DRL framework in Section~\ref{sec_drl} and its variants to improve the stability and learning performance. In Section~\ref{sec_mec}, as a concrete example, we shift our focus on performance optimization of the emerging MEC applications, which is generally complicated by the resource competition and interactions among multiple wireless users, base stations, caching and MEC servers~\cite{yan18}. We firstly build a general DRL framework to learn the optimal data offloading policy with uncertain network information, and then review the existing applications of DRL framework for MEC in different network scenarios. We observe that data offloading is not always preferred by low-power IoT devices due to the high energy consumption in wireless communications. Hence, in Section~\ref{sec_hybrid}, we introduce a novel hybrid MEC offloading model to balance the energy consumption in offloading and computation. Besides local computation, the hybrid model allows data offloading to the MEC server via either the active RF communications or the passive wireless backscatter~\cite{ieeenetwork}. Our numerical results verify that the hybrid MEC offloading can significantly improve the network performance, by learning the optimal transmission scheduling and workload allocation among different offloading schemes. Finally, some open issues are discussed in Section~\ref{sec_open}.

\section{An Overview of Deep Reinforcement Learning}\label{sec_drl}

In this section, we first review fundamentals of reinforcement learning and then discuss its extension to DRL, as well as various techniques to improve the learning efficiency and stability.

\subsection{Fundamentals of Reinforcement Learning}
Reinforcement learning is an effective solution to Markov Decision Processes (MDPs), which is composed of the decision-making agent, system state, action, and reward~\cite{sutton1998reinforcement}. The agent is the entity of decision making through interactions with the environment. Based on the observation of the environment, referred to as the system state, the agent takes an action and then receives an immediate reward correspondingly to the state-action pair. The action affects the environment and may cause the transition to a new system state. The immediate reward and the transition to a new state will guide, i.e., reinforce, the adaptation of the agent's policy, which defines the sequence of actions taken in each decision epoch as the system evolves. This learning process continues as we find the optimal policy to maximize the accumulated reward, which can be characterized by either the state-value or action-value function. The state-value records the expected total reward starting from an initial system state, while the action-value, also referred to as the $Q$-value, maps each state-action pair to the accumulated reward.

There are mainly value- and policy-based approaches for solving reinforcement learning problems~\cite{sutton1998reinforcement}. The value-based approach estimates the value function and takes the action to improve it directly in an iterative process. The estimation of value functions can be based on value iteration following the Bellman equation or ${Q}$-learning algorithm. A variant of the value-based approach relies on the estimate of an advantage-value, which can stabilize the learning process by subtracting a baseline from the estimate of action-value. The policy-based approach improves the value function by updating a parametric policy in gradient-based methods. Reinforcement learning can be also classified into on- and off-policy approaches. The on-policy learning relies on the sample trajectory induced by the current policy, i.e., all future actions are chosen according to the current policy. This may require more interactions with the environment to ensure unbiased policy updates and thus make it impractical for solving complicated problems. The off-policy learning can improve the sample efficiency by utilizing all historical sample trajectories. However, it requires more effort in hyperparameter tuning to ensure the convergence in learning.

\subsection{Deep Reinforcement Learning Approaches}

The reinforcement learning becomes unstable and even fail to converge when the state and action spaces are large in complex wireless networks. DRL can use the DNNs as function approximators for different components of reinforcement learning, including the value function, policy, and the underlying system model, e.g., the state transition probability. In the following, we introduce the basics of DRL and recent advances to improve its learning performance.

\subsubsection{Deep $Q$-Network (DQN)} It extends the value-based $Q$-learning algorithm for MDPs by using DNN as a parametric approximation for the action-value function~\cite{nature15}. The success of DQN and its variants relies on two key mechanisms, i.e., experience replay and target $Q$-network, to stabilize learning with large state and action spaces. The experience replay maintains a replay memory to buffer historical transition samples and randomly selects a subset of samples, i.e., mini-batch, to train the DNN. This can break the sample correlations and ensure more efficient training by independent samples. The training of DNN in each step aims to minimize the temporal-difference (TD) error, i.e., the mean-squared difference between the estimated $Q$-value by the DNN and its target value. Practically, we can replay more frequently the transition samples that generate a higher expected reward. Hence, a prioritized experience replay (PER) scheme can potentially increase the learning speed~\cite{per}. A straightforward way for PER is to prioritize samples by their TD-errors. A higher TD-error implies a larger potential to be further optimized. The TD-error based PER can be further combined with random sampling to ensure that all transition samples have the chance to be selected for training~\cite{sutton1998reinforcement}.

\begin{figure}[t]
  \centering
  \subfloat[DQN]{\includegraphics[width=0.33\textwidth]{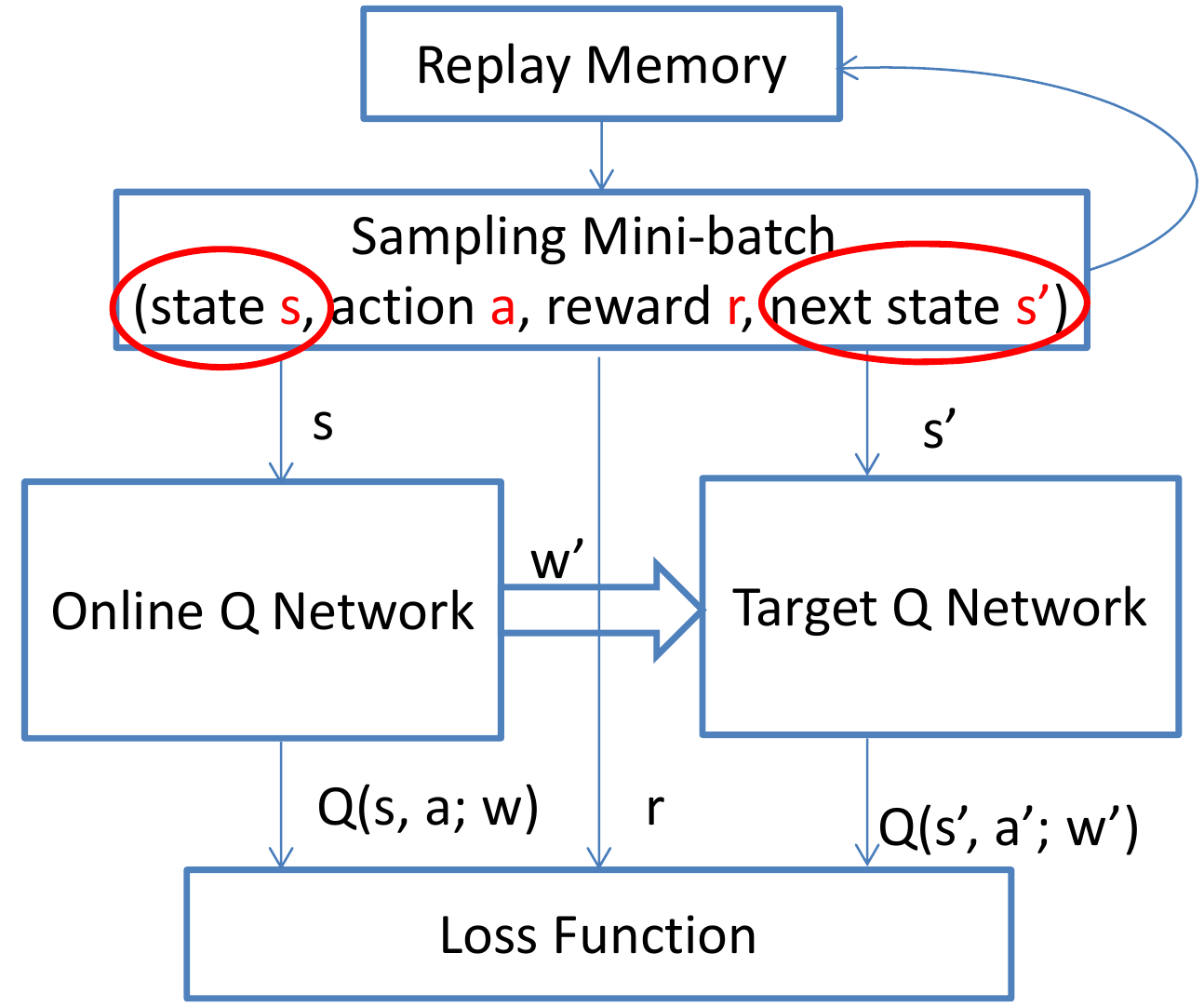}}
  \subfloat[DDQN]{\includegraphics[width=0.33\textwidth]{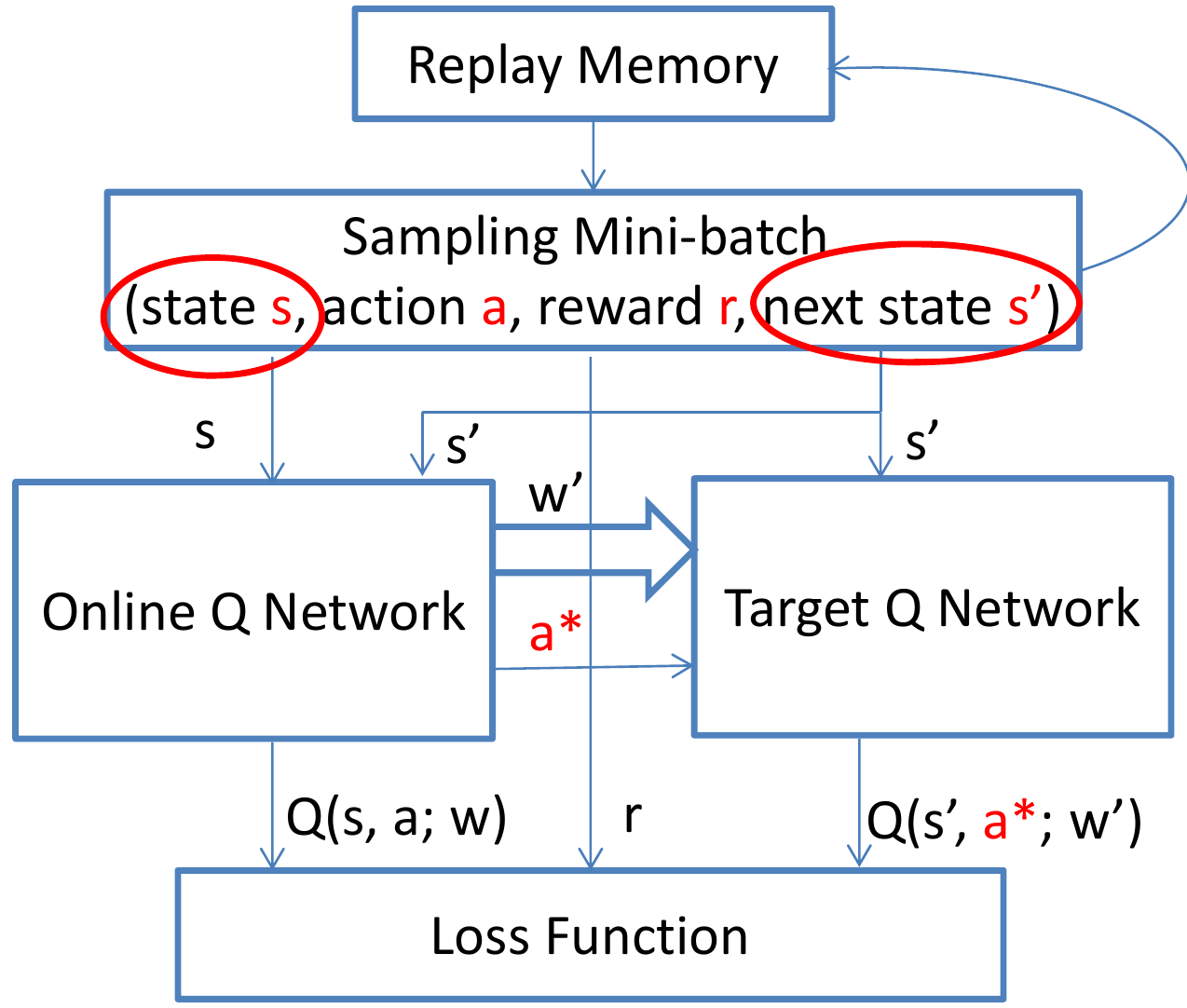}}
  \subfloat[Dueling DDQN]{\includegraphics[width=0.33\textwidth]{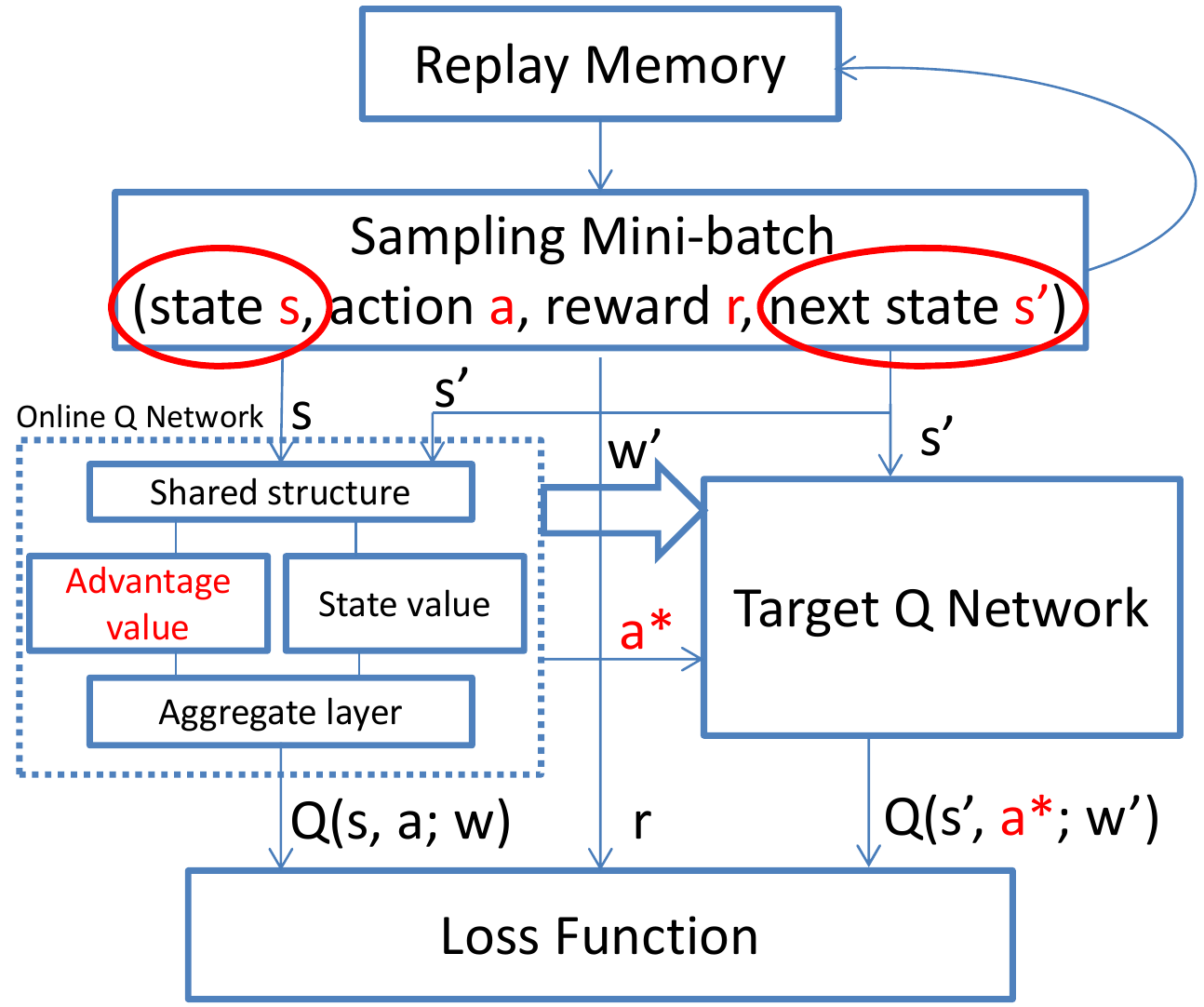}}\\
  \subfloat[Comparison of different DRL approaches.]{
  \resizebox{\textwidth}{!}{\begin{tabular}{|l|l|l|l|l|l|l|}\hline
  DRL approaches & Year &  Action/State & DNN setting & Learning speed & Sample efficiency & On/Off-policy \\\hline
  Plain DQN~\cite{nature15} & 2015 & Disc. moderate & One DNN& Slow & Low  & Off-policy  \\\hline
  DQN with PER~\cite{per} & 2016 & Disc. moderate & One DNN& Moderate & High  & Off-policy \\\hline
  DDQN with PER~\cite{ddqn} & 2016 & Disc. large  & Two DNNs & Fast & High & Off-policy  \\\hline
  Dueling DDQN~\cite{dueling} & 2016 & Disc. large & Three DNNs& Fast & High  & Off-policy  \\\hline
  DDPG~\cite{ddpg} & 2016 & Cont. large & Two DNNs & Fast & Moderate  & Off-policy   \\\hline
  Rainbow~\cite{rainbow} & 2018 & Disc. large & Adaptive & Faster  & High   & Mixed \\\hline
  TRPO~\cite{on_trpo} & 2019 & Cont. large & Two DNNs & Fast & Moderate  & On-policy   \\\hline
  \end{tabular}}
  }\caption{The comparison of typical value-based and policy-based DRL approaches.}\label{fig_ddqn}
  \vspace{-0.5cm}
\end{figure}

\subsubsection{Double and Dueling DQN} DQN uses a separate $Q$-network to generate the target value. The target $Q$-network updates its parameter by copying it from the online $Q$-network in every a few steps, as illustrated in Fig.~\ref{fig_ddqn}(a). This can make the learning more stable compared to the $Q$-learning algorithm. The drawback of DQN lies in that it uses the $\epsilon$-greedy policy to select an action and evaluate it by the same $Q$-network~\cite{sutton1998reinforcement}. This may lead to over-optimistic estimation of the $Q$-value. To correct this, Double DQN (DDQN) updates the action by the online $Q$-network and then evaluates it by the target $Q$-network~\cite{ddqn}, as illustrated in Fig.~\ref{fig_ddqn}(b). Another variant of DQN decomposes the $Q$-value into two streams, i.e., the state-value and the advantage-value~\cite{dueling}, approximated by two independent DNNs in a dueling architecture. The two streams are then combined via an aggregating layer to produce the final estimate of the $Q$-value. 

\subsubsection{Deep Deterministic Policy Gradient}

The policy-based and value-based approaches can be combined in the actor-critic framework~\cite{sutton1998reinforcement}. The critic function produces the estimation of the $Q$-value by minimizing the TD-error. The actor function then updates the policy parameter using the critic's feedback. Two independent DNNs can be used as the parametric approximations for the critic and actor functions, respectively. The intuition behind actor-critic framework stems from the policy gradient theorem that builds the connection between policy gradient and the $Q$-value. It decomposes the gradient computation into the evaluation of the $Q$-value and the gradient of the parametric policy, averaged over the whole state and action spaces. One recent development is to extend the policy gradient theorem to deterministic policy gradient (DPG), which outputs a deterministic action instead of a distribution on action space and thus makes it more efficient to estimate the policy gradient. The deep deterministic policy gradient (DDPG) algorithm combines DQN and DPG in the actor-critic framework to make the learning more stable and robust by using the experience replay and target $Q$-network for DNN training~\cite{ddpg}.

A comparison of typical DRL approaches is listed in Fig.~\ref{fig_ddqn}. In general, DQN and its variants are applicable to discrete action space, which are natural extensions of $Q$-learning algorithm for solving MDPs with large action and state spaces. The Rainbow algorithm in~\cite{rainbow} is an integrated design of different DQN variants, which achieves the best learning speed and maximum reward. The continuous action space can be more preferably tackled by DDPG in~\cite{ddpg} and the trust-region policy optimization (TRPO) in~\cite{on_trpo}. To avoid large variance in gradient estimation, TRPO formulates a constrained optimization to search for a better policy that improves the value function. Besides, we observe that the off-policy is more popular for DRL as it can use all historical samples efficiently. Though TRPO is generally on-policy, it has been adapted in~\cite{on_trpo} to leverage a replay buffer and thus can achieve a better learning performance compared to DDPG.

As modern wireless networks become large-scale and complicated, the network control problems face very diversified decision variables, including both discrete indicators and continuous variables for resource allocation. Thus, both value- and policy-based methods need to be used jointly for mixed decision-making problems. In the following, we focus on the applications of DRL in the emerging MEC offloading scenarios, which typically involve the interactions among multiple network entities and complicated optimization in both discrete and continuous domains.

\section{DRL-based Data Offloading for Mobile Edge Computing}\label{sec_mec}

MEC offloading allows IoT devices to offload data and computation-intensive workload (e.g., compressing and encryption) to resource-rich MEC servers. It can potentially reduce the processing delay, extend the battery lifetime, and even enhance security for IoT applications~\cite{yan18}. One of the critical design issues is to optimize the offloading rate, workload allocation, and choose the optimal MEC server, considering the time-varying channel conditions, user mobility, energy supply, dynamic workloads, and various resource constraints. A joint optimization on caching, offloading, networking, and transmission control is usually very complicated due to close couplings among multiple wireless users, base stations, and MEC servers. The optimization is also very inflexible to capture the network dynamics with uncertain parameters, e.g., the fluctuating channel conditions, the time-varying workload and energy supplies.

\subsection{General DRL Framework for MEC Offloading}\label{sec_hybrid}

\begin{figure}[t]
  \centering
  \includegraphics[width=0.9\textwidth]{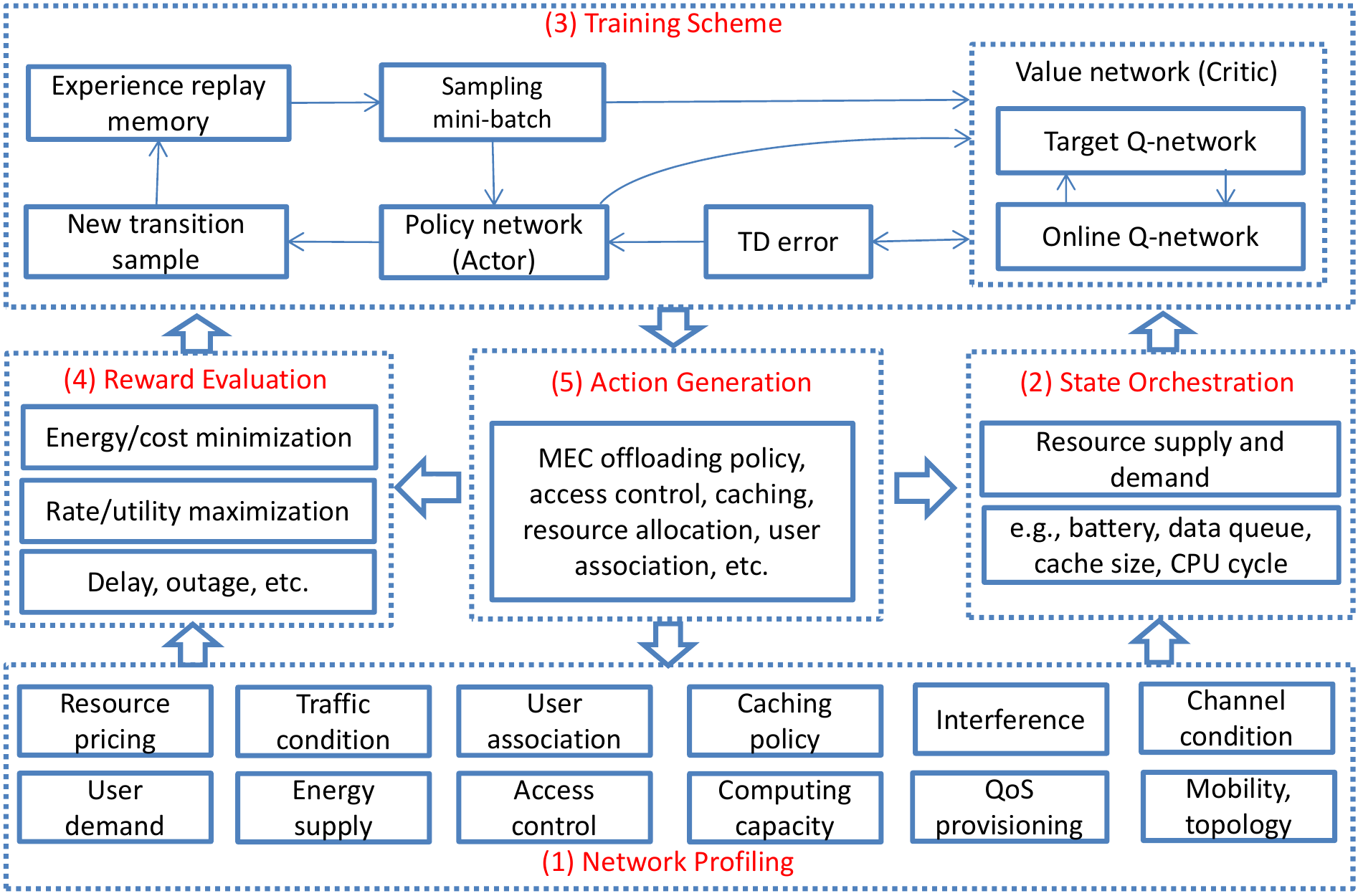}
  \caption{DRL-based MEC offloading framework.}\label{fig_mecdrl}
  \vspace{-0.5cm}
\end{figure}

DRL avoids above-mentioned difficulties by reformulating the network control problem into the MDP framework and enhancing the reinforcement learning solution by deep learning techniques. In the sequel, we propose a general DRL framework for MEC offloading that can be flexibly tailored for learning offloading strategy under different network scenarios. As illustrated in Fig.~\ref{fig_mecdrl}, the DRL framework includes the following main components: \begin{itemize}
  \item[(1)] Network Profiling: The network environment contains very high dimensional information. Dimension reduction is required to speed up the learning process. Network profiling helps to extract problem-dependent information closely related to the network control problems. This can be assisted by conventional model-based optimization problems.
  \item[(2)] State Orchestration: It aims to select the most salient or indicative state variables to minimize the state space without compromising the learning performance. The network performance depends on the demands and supplies of various resources. Hence, the system state can be set to show the real-time dynamics of resource consumption and its regeneration.
  \item[(3)] Training Scheme: The training scheme can flexibly organize the value and policy networks to learn both discrete and continuous offloading decisions, e.g., the discrete indicators for base station (de)activation, channel assignment, user association, and routing selection, as well as the continuous variables for bandwidth allocation and beamforming optimization.
  \item[(4)] Reward Evaluation: The reward in each decision epoch drives the DRL agent to adjust its MEC offloading policy. Practically, the reward is evaluated after completing the workload after a few decision epoches or time slots. A model-based optimization can be deployed to estimate the instant reward based on the prediction of future network dynamics.
  \item[(5)] Action Generation: The DRL agent outputs a vector of actions for each system state, which will be translated into the control variables to execute the offloading decisions. Quantization or approximation can be required in some cases to project continuous variables into discrete actions. Random noise can be also added to continuous actions for a better exploration.
\end{itemize}

The general DRL framework can be applied to optimize MEC offloading policies under different network scenarios by customizing different components of the DRL framework to meet the performance requirements of various design problems. In the following, we provide a review of recent works on the applications of DRL for MEC offloading problems.

\subsection{Design Issues for DRL-based MEC Offloading}\label{sec_rfpowered}

\subsubsection{Network Selection for Cost Minimization} In the simplest case with one wireless user and multiple access points, e.g., cellular base station and WLAN access point~\cite{zhangdeep}, the MEC offloading is regarded as a network selection problem as illustrated in case (i) of Fig.~\ref{fig_ddqn}. The wireless user can either access the cellular network or WLAN with different costs. To minimize the user's energy consumption and cost for channel access, DQN can be constructed to learn the optimal selection scheme without knowing the user's mobility pattern. The offloading decision is made based on the prediction of the user's location and the remaining data size.

\subsubsection{Channel and Capacity Sharing} When multiple wireless users request for the computation resources simultaneously from a single MEC server, e.g.,~\cite{xdwang_ddpg18}, as shown in case (ii) of Fig.~\ref{fig_ddqn}, the spectrum and capacity sharing becomes a critical problem to minimize the cost of delay and power consumptions for all users. The system state can be the sum cost of all users and the remaining capacity of the MEC server. The DRL agent learns the continuous resource allocation for wireless users and the binary offloading decisions, considering a limited capacity of the MEC server and time-varying channel conditions.

\begin{figure}[t]
  \centering
  \subfloat[Applications of DRL-based MEC offloading framework in different network scenarios: (i) Network selection~\cite{zhangdeep}, (ii) Channel and capacity sharing~\cite{xdwang_ddpg18}, (iii) MEC server and user association~\cite{chen2018optimized}, (iv) Collaborative data offloading~\cite{duc2018deep}.]{\includegraphics[width=0.99\textwidth]{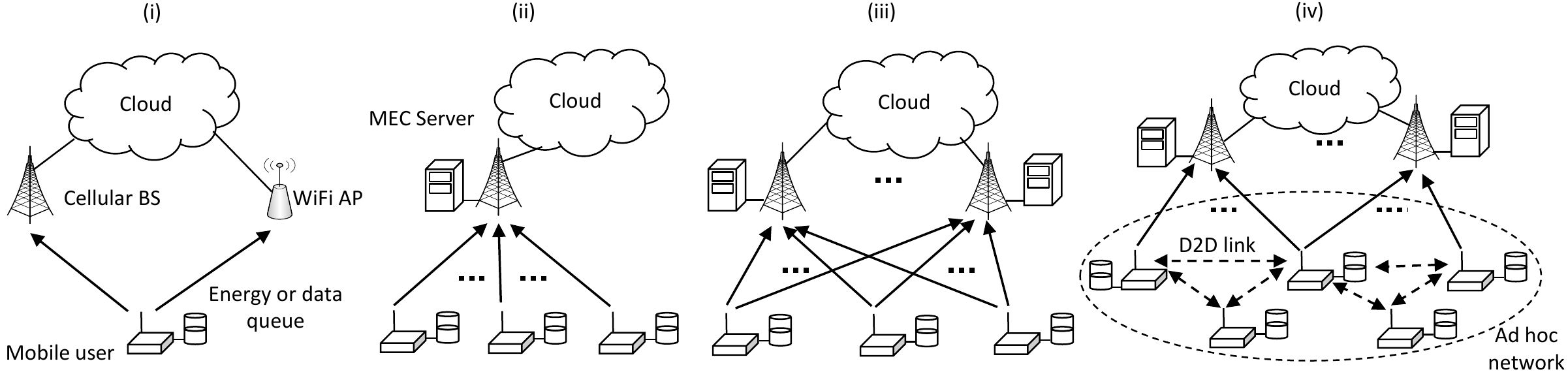}}\\
  \subfloat[The states, actions, rewards definitions in different DRL approaches for MEC offloading.]{
  \resizebox{\textwidth}{!}{
  \begin{tabular}{|m{1.5cm}|m{2.5cm}|m{4cm}|m{5cm}|m{5cm}|}\hline
  Scenarios & DRL models &  States & Actions & Rewards \\\hline
  (i)~\cite{zhangdeep} & \makecell[l]{DQN \\ using CNN} & {User's location and the remaining file size} & {Keep idle, access WLAN or cellular network} & {Minimize cost and energy consumption} \\\hline
  (ii)~\cite{xdwang_ddpg18}& DDPG & {Channel states and all users' task buffers}  & {Power distribution in local and offloading computation} & Minimize energy consumption within delay deadline\\\hline
  (iii)~\cite{chen2018optimized} & Double DQN & Channel states, energy and task queues & User association and power allocation & Maximize satisfaction on delay, outage probability, and payment for MEC service \\\hline
  (iv)~\cite{duc2018deep} & Plain DQN & Task queues of all users and their interdistance & Task distribution in local and offloading computing  & Maximize utility minus the cost of energy consumption, delay, and outage probability \\\hline
  \end{tabular}
  }
  }\caption{DRL applications in different MEC offloading scenarios.}\label{fig_ddqn}
  \vspace{-0.5cm}
\end{figure}

\subsubsection{MEC Server and User Association} With multiple base stations or MEC servers, each wireless user's computation offloading can be routed via different base stations, as shown in case (iii) of Fig.~\ref{fig_ddqn}. To minimize the cost of processing delay, the authors in~\cite{chen2018optimized} employed DDQN to learn the optimal offloading policy including the binary user association and transmit control strategies. The system state consists of the channel conditions between the wireless users and the base stations, the statuses of energy and data queues. Considering low utilization of base stations, DDQN can be also used to control the (de)activations of base stations to reduce total energy consumption while maintaining the same quality provisioning.

\subsubsection{Collaborative Data Offloading}
Besides offloading to the MEC server, the collaborative offloading among multiple wireless users can be envisioned in case (iv) of Fig.~\ref{fig_ddqn}, i.e., each wireless user can offload its computation workload to nearby users via device-to-device communications, e.g.,~\cite{duc2018deep}. The optimization of offloading decisions depends on the number of remaining tasks at each wireless user, the availability of the computation resources, and the link quality between wireless users. DQN or DDQN can be customized to learn the optimal offloading policy in a mobile ad-hoc network to maximize the resource utilization or minimize total power consumption, subject to the user's energy and delay requirements.


\section{DRL Approach for a Hybrid MEC Offloading Model}\label{sec_hybrid}

One design objective of the future wireless network is to embrace the ubiquitous interconnections of low-power IoT devices, e.g.,~the wearable wireless sensors for healthcare monitoring, either battery-powered or wireless powered via energy harvesting~\cite{xdwang_ddpg18}. For these low-power IoT devices, it is clear that energy consumption on data processing can be reduced significantly by offloading computation-intensive workload to the MEC servers, e.g.,~\cite{zhangdeep,xdwang_ddpg18,chen2018optimized,duc2018deep}. However, in another aspect, the energy saving on computation comes with the price of more energy consumption on computation offloading, which is generally performed by conventional RF communications. Due to the high energy consumption of RF communications, MEC offloading may not be affordable by these low-power IoT devices.

In this section, we tackle this problem by proposing a hybrid offloading strategy that can schedule data offloading in both high-rate RF communications and low-power backscatter communications~\cite{ieeenetwork}. The backscatter radio operates in the {passive} mode by reflecting the incident RF signals. It is featured with extremely low power consumption and a low data rate, while the active radio in RF communications can transmit reliably with a higher data rate by adapting its transmissions against the channel fading effect. Hence, we aim to optimize the hybrid MEC offloading policy to balance energy consumptions in both data offloading and computation. This can be achieved by exploiting the complement operations of the passive and active radios. However, due to the couplings among two radio modes, it becomes more complicated to optimize the MEC offloading policy by using the conventional model-based approaches. In the sequel, we employ the DRL framework to optimize the hybrid MEC offloading strategy with uncertain channel conditions, dynamic energy supply, and time-varying workloads at the IoT devices.

\begin{figure}[t]
  \centering
  \includegraphics[width=0.9\textwidth]{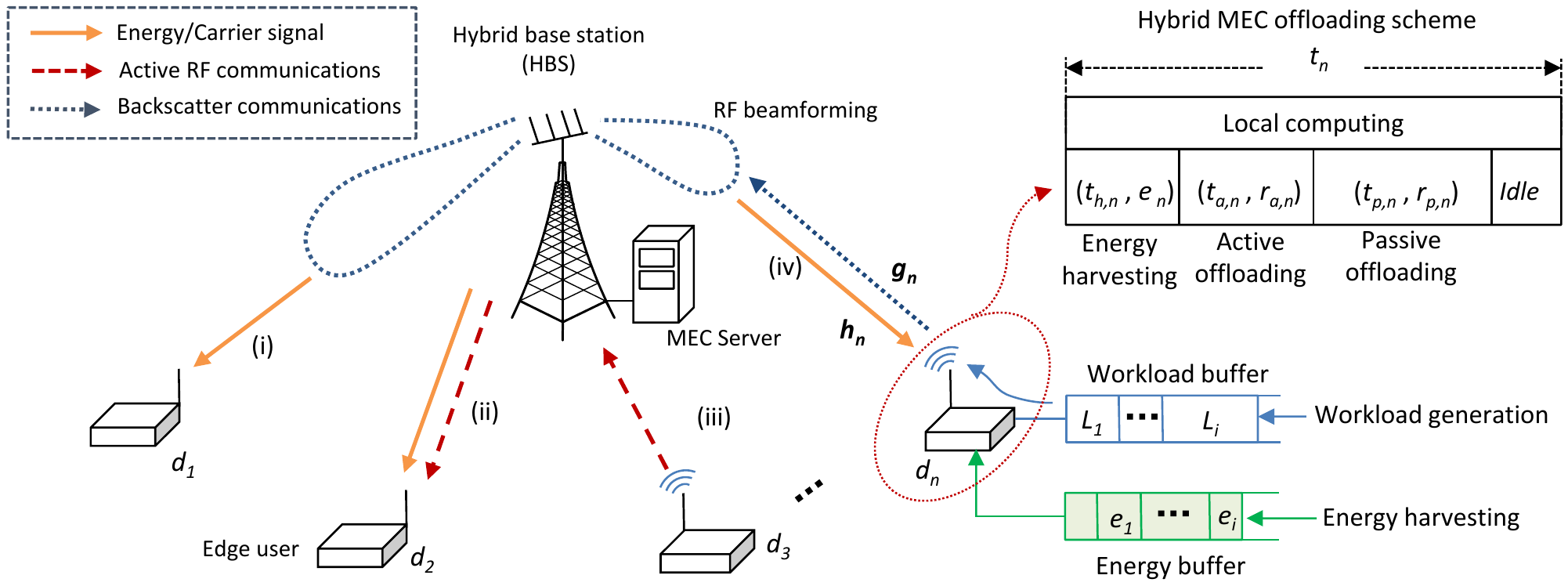}
  \caption{Hybrid MEC offloading scheme: (i) RF-based energy harvesting, (ii) Active offloading via RF communications, (iii) Down-link power and information transfer, (iv) Passive offloading via backscatter communications.}\label{fig_hybridmec}
  \vspace{-0.5cm}
\end{figure}

\subsection{Hybrid MEC Offloading Model}
We consider a set of edge users, e.g., wireless IoT sensors, that send backlogged workloads to a hybrid base station (HBS), which is co-located with the MEC server. The system model is illustrated in Fig.~\ref{fig_hybridmec}. The channel from the HBS to each edge user is modeled by a finite-state Markov model. The HBS allocates each edge user a time slot for MEC offloading, similar to the time-slotted structure in~\cite{xdwang_ddpg18}. Each edge user can harvest RF energy from the HBS and the ambient RF sources with random power density. The energy harvesting capability is illustrated in case (i) of Fig.~\ref{fig_hybridmec}. The edge user's workload is uncertain due to the user's mobility and time-varying demand of upper layer applications, e.g., the data sampling rate may vary according to the health conditions of the subject being monitored. The user's workload needs to be processed locally or remotely at the MEC server before a time deadline. We assume that the MEC server can return the processed data to the edge user instantly via simultaneous power and information transfer, as illustrated in case (ii) of Fig.~\ref{fig_hybridmec}. The hybrid MEC offloading scheme allows each user to flexibly switch data offloading between the passive backscatter communications and the active RF communications, as illustrated in cases (iii) and (iv) of Fig.~\ref{fig_hybridmec}, respectively. To maintain a fixed offloading rate, the active radio's transmit power has to be adapted with the time-varying channel conditions. This implies a dynamic process of the edge user's energy buffer.

It is obvious that the offloading scheduling between two radio modes introduces an additional degree of freedom to improve the MEC performance in such a dynamic network environment. The DRL approach aims to learn the optimal hybrid MEC offloading policy from past experience. Given the channel conditions, energy status, and workload in each time slot, the edge user will choose its action (e.g., local computation, passive or active offloading) to maximize the reward function, which is defined as the energy efficiency, i.e., the successfully processed workload per unit energy. Workload outage happens when the edge user's workload is not processed successfully before the delay bound. In this case, the instant reward will be set to zero. To proceed, we divide each time slot into flexible sub-slots as illustrated in Fig.~\ref{fig_hybridmec}. The first sub-slot $t_{h,n}$ is reserved for RF energy harvesting. The following sub-slot $t_{a,n}$ is allocated to active offloading with a higher rate $r_{a,n}$ and another sub-slot $t_{p,n}$ is used by passive offloading with a lower rate $r_{p,n}$. The offloading schemes also differ in their power consumption. To achieve the maximum energy efficiency, the DRL agent is designed to learn a transmission scheduling policy that determines the optimal action on each system state, including time and workload allocations among local computation, active and passive offloading, subject to the edge user's energy budget constraint.

\subsection{Numerical Evaluation}

\begin{figure}[t]
  \centering
  \subfloat[Maximum reward with the Hybrid-Offload scheme]{\includegraphics[width=0.5\textwidth]{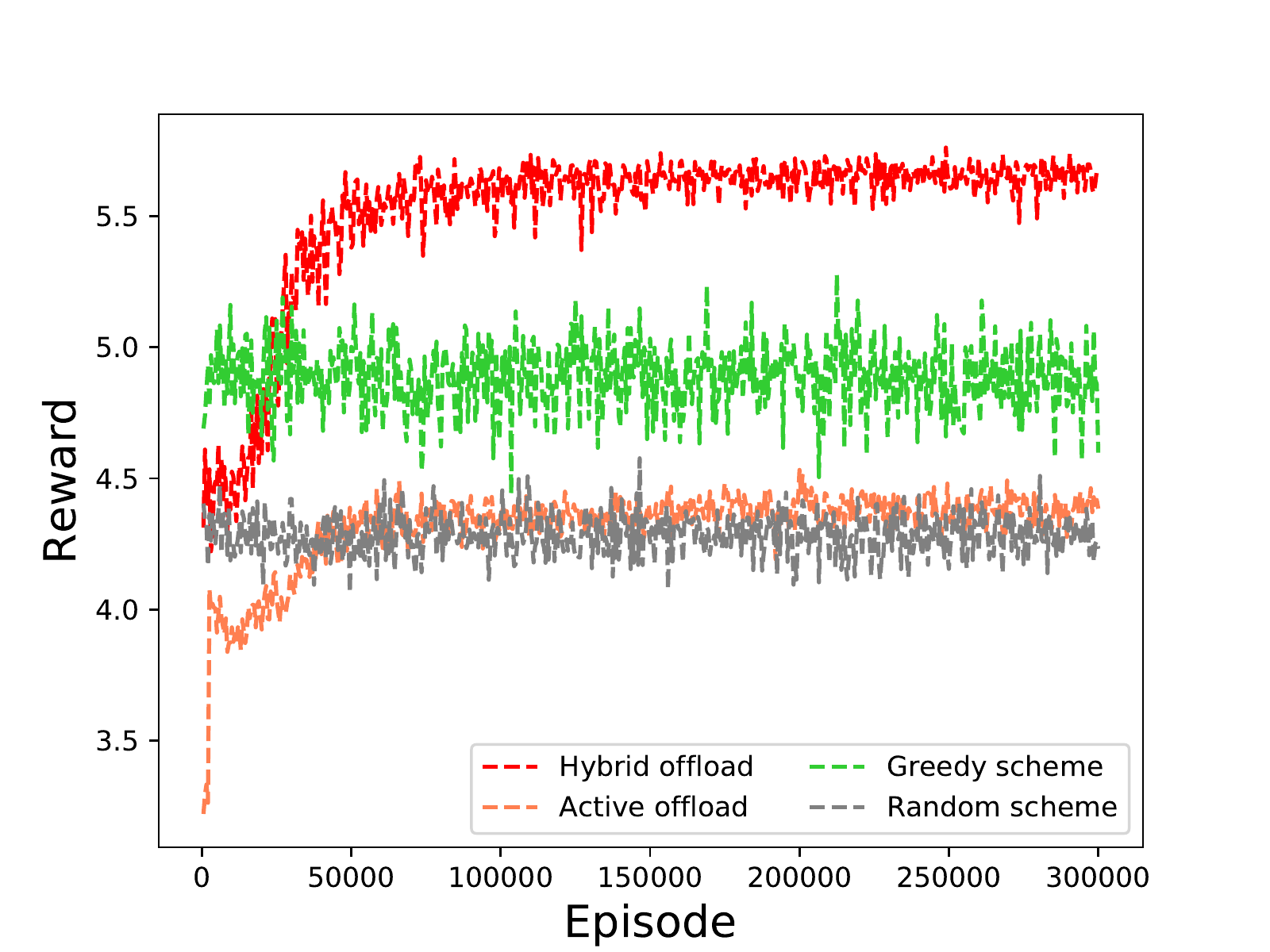}}
  \subfloat[Minimum outage with the Hybrid-Offload scheme]{\includegraphics[width=0.5\textwidth]{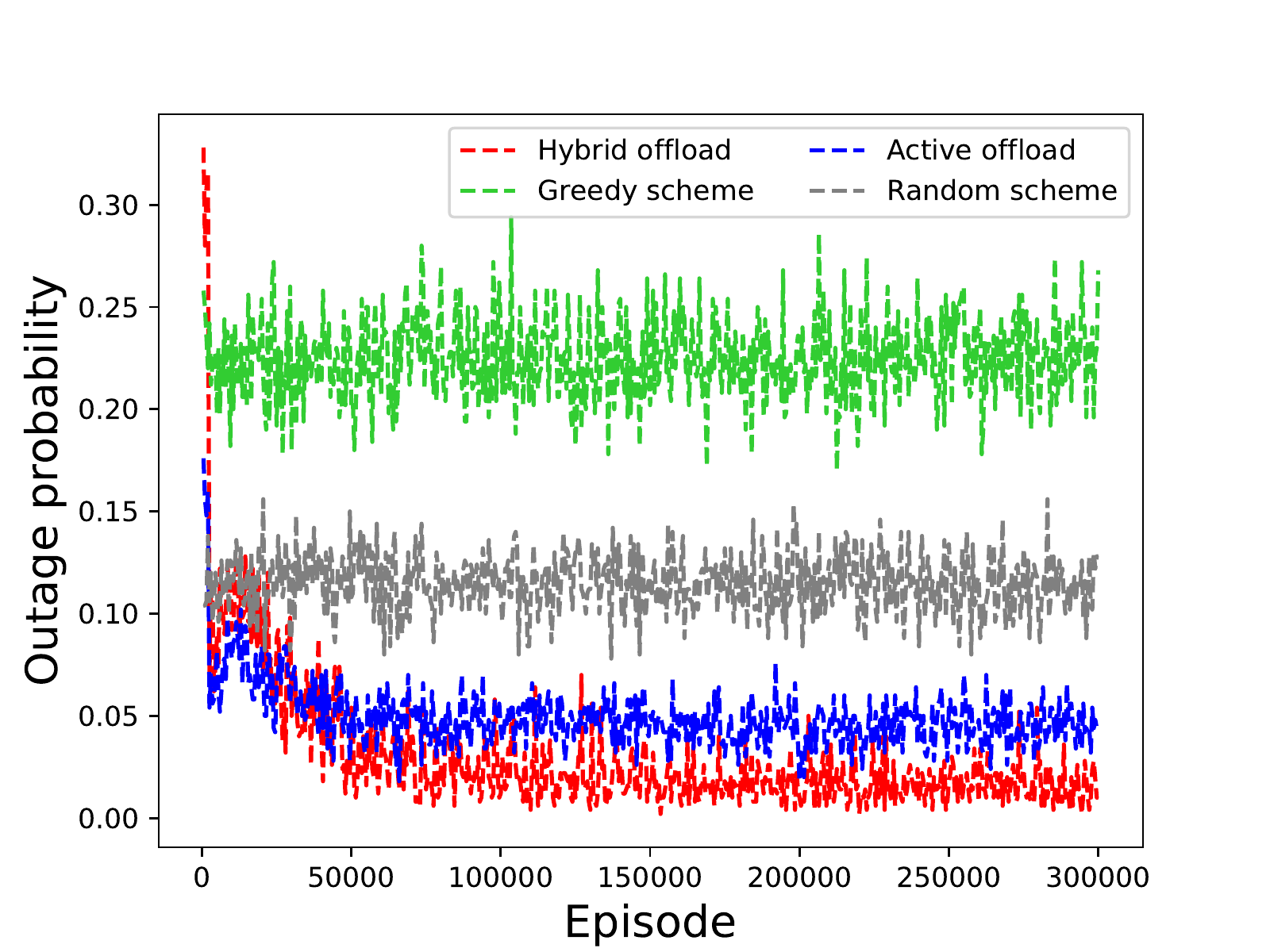}}
  \caption{Performance comparison among different MEC offloading schemes.}\label{fig_diff_schemes}
  \vspace{-0.5cm}
\end{figure}

To exploit the flexibility and performance gain via hybrid MEC offloading scheme, we compare it with the conventional offloading scheme, namely, the Active-Offload scheme that only supports active RF communications, e.g.,~\cite{xdwang_ddpg18}. We also compare it with the typical greedy and random schemes. In Fig.~\ref{fig_diff_schemes}, we show the performance of different schemes and observe that the DRL-based Hybrid-Offload scheme achieves the highest reward and the lowest outage performance, as shown in Fig.~\ref{fig_diff_schemes}(a) and Fig.~\ref{fig_diff_schemes}(b), respectively. The Active-Offload scheme uses a similar DRL framework as that of the Hybrid-Offload scheme, however with the reduced action space, i.e.,~it only chooses between local computation and active offloading. Hence, it achieves a reduced reward performance than that of the Hybrid-Offload scheme. The benchmark greedy scheme always chooses the myopic action to maximize the instant reward in each time slot. It even performs better than the Active-Offload scheme due to its flexibility in radio mode switching.

\begin{figure}[t]
  \centering
  \subfloat[Higher reward can be achieved with more sub-slots]{\includegraphics[width=0.5\textwidth]{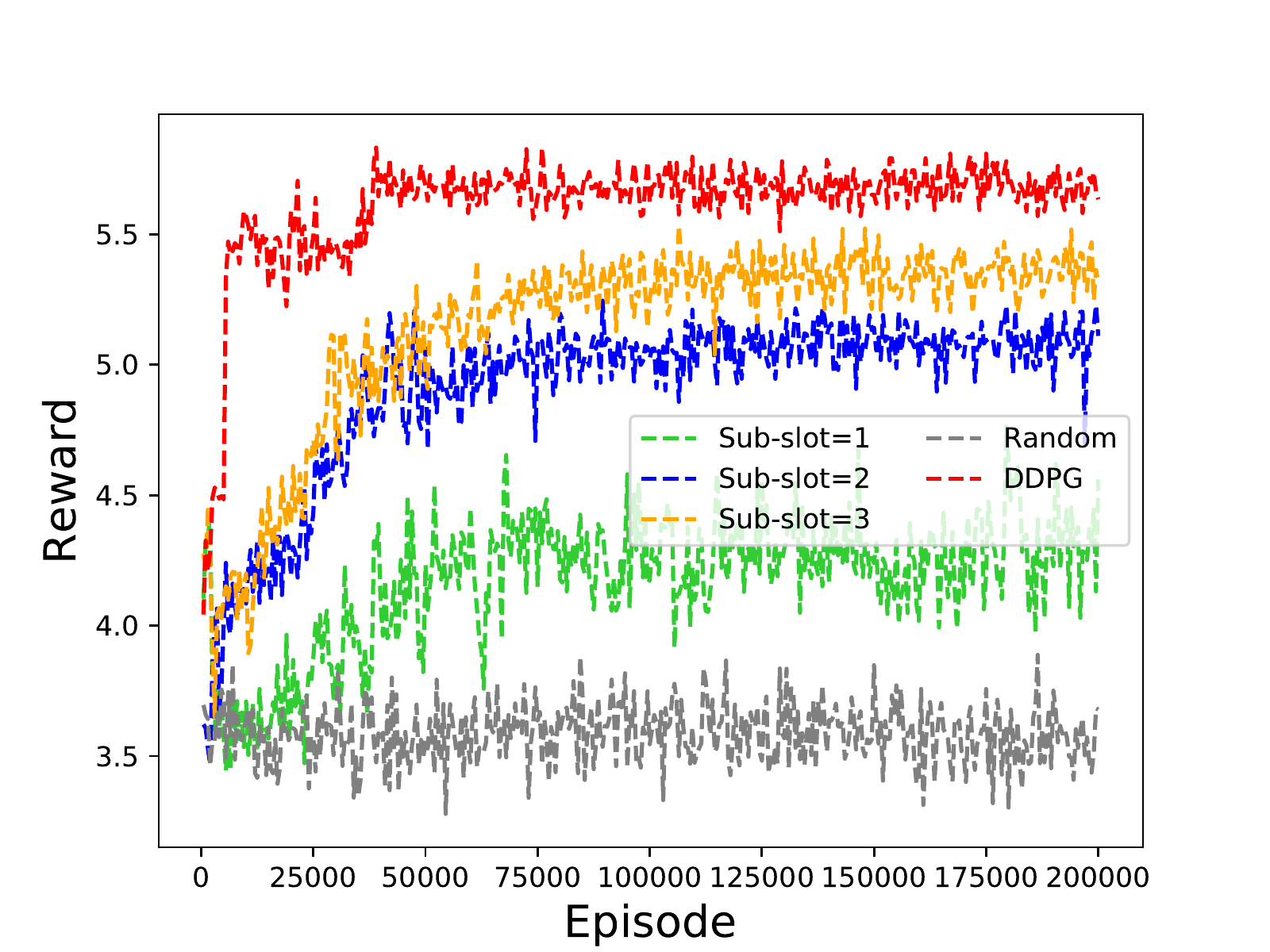}}
  \subfloat[Workload allocation among three computation schemes]{\includegraphics[width=0.5\textwidth]{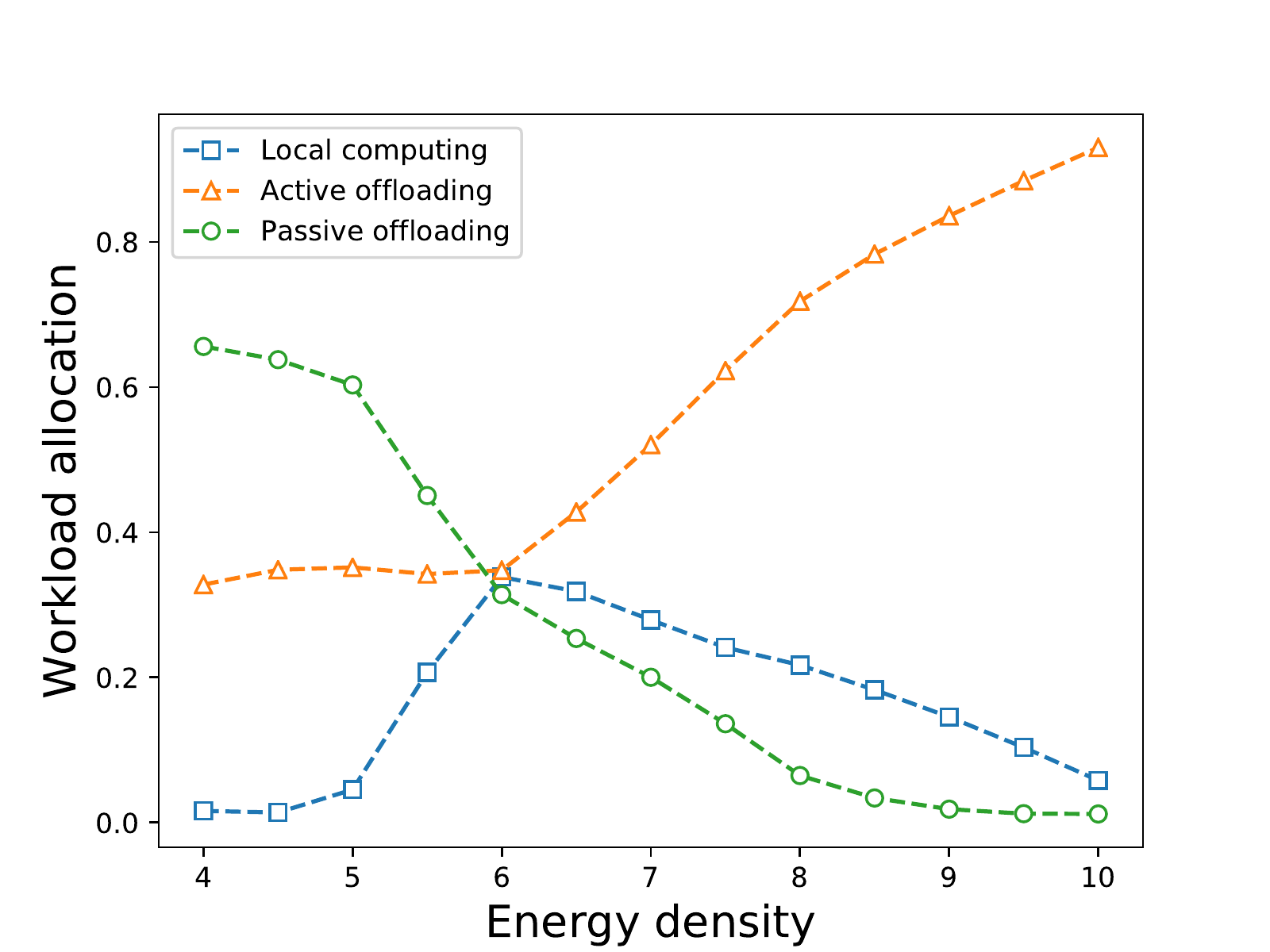}}
  \caption{Performance comparison with different parameter settings.}\label{fig_diff_setting}
  \vspace{-0.5cm}
\end{figure}
In the next group of simulations, we equally divide each time slot into multiple sub-slots and assume that the edge user follows the same DRL framework to optimize its offloading decision independently in each sub-slot. By this way, we can flexibly allocate the workload and thus approximate the optimal workload allocation strategy among local computation, passive and active offloading. In Fig.~\ref{fig_diff_setting}(a), we show the performance of the Hybrid-Offload scheme when we set a different number of sub-slots for MEC offloading. We observe that it generally achieves a higher reward performance with more sub-slots. We also show the performance of the DDPG algorithm for continuous control that is shown to achieve the maximum reward. Such a performance gain is obtained from the increased flexibility in workload and time allocation.

In Fig.~\ref{fig_diff_setting}(b), we show the averaged workload allocation among different computation schemes at the convergence of the Hybrid-Offload algorithm. The $x$-axis of Fig.~\ref{fig_diff_setting}(b) denotes the mean power density in the ambient RF environment. We can observe that with low energy supply the passive offloading scheme is preferred due to its extremely low power consumption. With a higher energy density, the edge user generally harvests more RF energy and thus it prefers to use the active offloading scheme. This can provide a higher offloading rate and thus reduce the processing delay.

\section{Open Research Issues}\label{sec_open}

Though DRL has been successfully applied to various network control problems, there still exist some challenges and open issues for MEC offloading in wireless networks.

\subsubsection{Multi-agent DRL for MEC Offloading} MEC offloading involves multiple heterogenous network entities, e.g., wireless users, base station, and MEC server, which may have totally different reward functions and control variables. Each user can customize its own DRL framework and make decisions based on local observations. However, this may destroy the Markovian property of the underlying system model and lead to divergent learning performance.

\subsubsection{Model-based Reward Evaluation} The DRL agent requires real-time reward evaluation to drive the learning process. As the performance of MEC offloading decision is usually not observable until the completion of workload, we require a more effective way combining learning and model-based optimization to predict the reward with incomplete network information.

\subsubsection{Hierarchical DRL for MEC Offloading} MEC offloading decision generally involves both discrete and continuous control variables. To improve learning efficiency, a hierarchical or two-stage DRL framework can be implemented to learn the optimal resource allocation strategy by using the policy-based DRL approaches in the inner loop, and then update the discrete user association or offloading decisions by DQN or its variants in the outer loop.

\section{Conclusions}\label{sec_con}
In this paper, we firstly have reviewed the DRL framework for its applications in MEC offloading with uncertain network information. Then, we have customized the DRL framework to realize a novel hybrid MEC offloading scheme that exploits the complement transmissions of the passive and active radios. Numerical results demonstrate that it can significantly improve the offloading performance. In the last, we have outlined a few open research issues.

\bibliographystyle{IEEEtran}
\bibliography{drlreference}

\end{document}